\documentstyle[aasms4]{article}
\def\ltsima{$\; \buildrel < \over \sim \;$}
\def\simlt{\lower.5ex\hbox{\ltsima}}
\def\gtsima{$\; \buildrel > \over \sim \;$}
\def\simgt{\lower.5ex\hbox{\gtsima}}
\def\parsec{{\rm\,parsec}}

\def\s{\ifmmode \widetilde \else \~\fi}
\def\={\overline}

\def\spose#1{\hbox to 0pt{#1\hss}}

\def\lta{\mathrel{\spose{\lower 3pt\hbox{$\mathchar"218$}}
     \raise 2.0pt\hbox{$\mathchar"13C$}}}
\def\gta{\mathrel{\spose{\lower 3pt\hbox{$\mathchar"218$}}
     \raise 2.0pt\hbox{$\mathchar"13E$}}}
\def\Dt{\spose{\raise 1.5ex\hbox{\hskip3pt$\mathchar"201$}}}	
\def\dt{\spose{\raise 1.0ex\hbox{\hskip2pt$\mathchar"201$}}}	

\def\=={\equiv}

\def\dotsfill{\leaders\hbox to 1em{\hss.\hss}\hfill}

\newcommand{\ffffff}[1]{\mbox{$#1$}}
\newcommand{\scnd}{\mbox{\ffffff{''}\hskip-0.3em .}}
\newcommand{\scmd}{\mbox{\ffffff{''}}}

\newcommand{\micro} {micro--image}
\newcommand{\micros}{micro--images}
\newcommand{\macro} {macro--image}

\slugcomment{Submitted to the Astrophysical Journal}

\lefthead{G. F. Lewis \& R. A. Ibata}
\righthead{Quasar Image Shifts due to Gravitational Microlensing}

\begin{document}

\title{Quasar Image Shifts due to Gravitational Microlensing}

\author{Geraint F. Lewis} 
\affil{
Dept. of Physics and Astronomy, University of Victoria,
PO Box 3055, \nl Victoria, B.C.,  V8W 3P6, Canada \& \nl
Astronomy Dept., University of Washington, Box 351580, \nl
Seattle WA 98195-1580, U.S.A.
\nl
Electronic mail: {\tt gfl@astro.washington.edu}}

\medskip
\author{Rodrigo A. Ibata}
\affil{
European Southern Observatory, \nl
Karl-Schwarzschild Stra\ss e 2, \nl
D-85748 Garching bei M\"unchen, \nl
Germany.
\nl
Electronic mail: {\tt ribata@eso.org}}



\begin{abstract}
Gravitational microlensing of quasars by stars in external galaxies
can introduce fluctuations in the centroid of the ``point-like''
macro--images. The induced shifts are extremely small, on
micro--arcsecond scales, below the limits of current optical
observations. However, such shifts will become measurable with the
proposed ``Space Interferometry'' mission, due to fly in 2005. The
degree of the centroid shifts and their application as probes of both
quasar structure and the stellar mass function in the lensing galaxy
are discussed.
\end{abstract}




%

\newpage

\section{Introduction}\label{introduction}

Over the last decade gravitational microlensing has shown itself to be
not only a probe of the mass structure in our galactic halo (Alcock et
al. 1993), but also of the scales of structure in active galaxies
(e.g. Saust 1994, Lewis et al. 1997). Statistical studies of numerical
simulations have revealed that long term light curves of microlensed
quasars will additionally probe the mass function of the microlensing
objects (Lewis \& Irwin 1996).

In this paper, a complementary probe of these properties, via the
identification of microlensing induced ``centroid-shifting'' in the
images of macrolensed quasars, is presented.  The paper begins with a
description of gravitational microlensing due to stars in external
galaxies (Section~\ref{microlensing_section}). The microlensing model
is described in detail in Section~\ref{microlensingmodel_section},
with the results of this study being presented in
Section~\ref{results_section}.  These are discussed in the context of
space-borne interferometry in Section~\ref{space_section} and the
conclusions of this study are presented in the final section.

\section{Microlensing}\label{microlensing_section}

\subsection{Background}\label{microlensing_background}

Since the discovery of the first gravitational lens system (Walsh et
al. 1979), the action of the individual stellar mass objects on the
light curve of the macrolensed object has be considered (Chang \&
Refsdal 1979, Young 1981).  The problem introduces a natural
scale--length for an isolated star of mass ${\rm M}$, its Einstein
radius $\eta$, which is given by;
\begin{equation}
\eta = \sqrt{ \frac{\rm 4 G M }{\rm c^2 }
\frac{ \rm D_{os} D_{ls} }{ \rm D_{ol} } } ,
\label{ein_rad}
\end{equation}
where ${\rm D_{ij}}$ are angular diameter distances between the
observer (o), lens (l) and source (s)~(Schneider et al. 1992).  In
this case the length is defined in the ``source plane'' at a distance
${\rm D_{os}}$ from an observer.  A point-like source passing within
this projected radius of an isolated microlensing mass is amplified by
a factor of at least $1.34$, although the angular scale of multiple
imaging induced by stellar mass lenses is $\sim~10^{-6}{''}$, far
below present detection limits with optical telescopes. In practice,
therefore, observations of microlensed systems are confined to the
monitoring the image brightness fluctuations. The light curve induced
by the passage of an isolated microlens possesses a very simple form,
with a characteristic time-scale;
\begin{equation}
\tau \equiv \frac{ \eta }{\rm V_{eff} } ,
\label{timescale}
\end{equation}
where ${\rm V_{eff} }$ is the effective velocity of the source across
the source plane~(Kayser et al. 1986). In such simple cases, $\tau$ can
be used to characterize the mass of the microlenses~(Wambsganss 1992).
 
For a distant multiply--imaged quasar, whose light can shine through
the inner regions of the macrolensing galaxy, the optical depth to
microlensing can be substantial and the action of the individual
lensing masses combine in a highly non-linear fashion, with the light
curve of a background source exhibiting complex variability, including
strongly asymmetric fluctuations~(Paczy\'{n}ski 1986, Wambsganss et
al. 1990).  In such a regime, the time-scale of individual ``events''
no longer reflects the time-scale given by Equation~\ref{timescale};
such time--scales reflect the time taken for a caustic to sweep across
the quasar continuum source and, as the details of the caustic network
are due to the sum effect of the ensemble of microlenses, the
properties of the lensing bodies must be determined from statistical
considerations (eg Lewis \& Irwin 1995,1996).
 
Chang (1984) demonstrated that the amplification of a source is
dependent on its size relative to the Einstein radius of the
characteristic lensing mass. At caustic crossings a point source is
amplified by an infinite amount, but any physical extent leads to a
finite amplification. Sources whose scale size is only a fraction of an
Einstein radius can be amplified by large factors, while sources whose
scale is much greater than an Einstein radius are amplified by a
negligible amount.

\subsection{Q2237+0305}\label{microlensing_q22370305}

Discovered during the CfA redshift survey~(Huchra et al. 1985), the
Q2237+0305 system consists of four images of a background quasar,
$z\sim1.69$, separated by $\la~2\scnd0$~(Yee 1988). The images are
centered on the core of the lensing galaxy, a bright nearby,
$z\sim0.04$, barred spiral.  The galaxy light in the inner regions of
the Q2237+0305 system is dominated by stars in the central bulge of
the galaxy~(Yee 1988), and it is expected that optical depth to
microlensing is substantial.  This lensing galaxy is, however, unique
among known gravitational lens systems in that it is located very
close to the observer and thus the light--travel time differences to
the four images are only $\sim1\,$day.  Intrinsic variations in the
luminosity of the quasar thus manifest themselves in all four images
within $\sim1\,$day and differential variability between images with
time-scales $\gg1$day reflects the effects of microlensing. It is this
effective decoupling of the observed time-scales for intrinsic
variability and microlensing by stars in the galaxy that allowed
photometric monitoring of Q2237+0305 to produce the first detection of
a microlensing event~(Irwin et al. 1989, Corrigan et al. 1991).
Microlensing induced variability is a prominent feature of light
curves of almost a decades monitoring of this system ({\O}stensen et
al. 1996).

However, for Q2237+0305, the Einstein radius (Equation~\ref{ein_rad})
for a star of Solar mass, projected into the source plane, is $ \eta_0
\sim 0.04h_{75}^{\scriptscriptstyle{-\frac{1}{2}}} \parsec .  $
(Through--out we adopt a standard Freidmann-Walker Universe, with
$\Omega=1$ of matter distributed smoothly. The cosmological constant,
$\Lambda$, is assumed to be zero).  Within the framework of the
``standard model'' for the central regions of quasars~(Rees 1984) this
microlensing scale is smaller than the extent of the broad emission
line region, $0.1$--$1.0\parsec$, but significantly in excess of the
size of the ultraviolet--optical continuum-producing region, $\simlt
10^{-3}\parsec$. Thus, differential microlensing amplification of the
quasar continuum and emission line producing regions by objects of
stellar mass within the lensing galaxy of the Q2237+0305 system is
expected.  This is observable as a time--dependent variation in the
relative strengths of the broad emission lines and the continuum which
is coupled with microlensing--induced photometric variability (Lewis
et al. 1997).

\subsection{``Improper Motions''}\label{microlensing_microimages}

Williams and Saha (1995) addressed the degree of centroid shifting
during microlensing at substantial optical depth.  For this study, a
new model for mass distribution was constructed with non-parametric
techniques, recovering both the relative image positions and
magnitudes. Unlike previous models of this system, which predict a
single image at each of the image positions (with each image
consisting of a myriad of micro--images), the images in this model
comprise of several ``mini--images'', separated by $\sim~0\scnd05$.
These mini--images suffer differing convergence and shear and,
therefore, undergo differing amplifications. The close proximity of
these mini--images implies that they are not resolvable with current
optical telescopes~\footnote{Such image splittings may be accessible
through radio observations with the VLBI, which are capable of probing
milli--arcsec scales, although Q2237+0305 is not a powerful radio
source with observed flux densities of ${\rm 40-80~\mu~Jy}$ at 3.6~cm
(Falco et al. 1996).}, appearing as a single, point--like image whose
central position is the centroid of the light distributions of the
mini--images.

As the light from each of these mini--images shines through the
lensing galaxy it can be influenced by the action of microlensing
bodies. This causes the brightness of each mini--image, and therefore
the centroid of the light distribution, to vary over time. For the
quadruple lensing system, Q2237+0305, assuming the lensing galaxy
possesses a transverse velocity of ${\rm V = 600km~s^{-1}}$,
observations on time--scales of one year have a 50\% chance of
revealing shifts of $\simgt0\scnd003$, while on ten year time--scales
there is similar chance of $\simgt0\scnd01$, and a 15\% chance of
$\simgt0\scnd03$.  Centroiding of images to a fraction of a resolution
element, typically $\sim0\scnd01$, is now routine and Williams and
Saha (1995) propose that observed centroid shifts of these magnitudes,
observed in Q2237+0305, could be due to their proposed mechanism.

It should be noted, however, that the splitting of a macro--image into
several mini--images which experience differing microlensing
characteristics is particular to the modeling procedure of Williams
and Saha (1995); other techniques predict a single macroimage
comprised of micro--images which have a typical separation of
micro--arcsec.  In this paper we examine this regime by centroid
shifts introduced by the variation in brightness of these
micro--images as stellar field passes in front of the macrolensed
source.

\section{Microlensing Model}\label{microlensingmodel_section}

\subsection{Method}\label{microlensingmodel_method}

A number of techniques have been developed to numerically study the
action of microlensing objects on distant sources (Young 1981,
Paczy\'{n}ski 1986, Witt 1993, Lewis et al. 1993). The ``work--horse''
of numerical simulation has been the backwards ray--tracing method
(Kayser et al. 1986, Wambsganss et al. 1990) in which a regular grid
of rays is fired from an observer into a field of microlensing masses
lying in the ``image plane''. The gravitational lens action of these
masses act to deflect the ray.  The ray is then ``traced'' until it
impacts the source plane. Although the input rays are regularly
distributed, the non-linear action of the microlenses ensures the
distribution of rays over the source plane is not
(Figure~\ref{fig_microlens}) and the density of rays at a point in
the source plane, relative to the density of rays over the image
plane, is a measure of the amplification a source at that position
would experience.

To achieve sufficient resolution over the source plane, however, a
large number of rays needs to be traced through the field of
microlensing masses. Compounding this, the direct calculation of the
deflection angle suffered by a ray due to a large number of these
masses is computationally expensive and simulations involving direct
calculations would take a prohibitive time to complete. To alleviate
this, the gravitational microlensing ``force'' is approximated with
the use of tree codes (Barnes and Hut 1986), allowing realistic
simulations to be performed on desk--top computers.

Such a code is employed in this study. The source region was chosen to
be 2.5 Einstein Radii for a Solar mass, and pixelated to $256^2$. With
this, the area of source plane under consideration is
${\rm \sim3.2\times10^{17}h_{75}^{\scriptscriptstyle{-\frac{1}{2}}}}$cm, and
each source plane pixels is
${\rm \sim1.2\times10^{15}h_{75}^{\scriptscriptstyle{-\frac{1}{2}}}}$cm
along a side. A ray fired through the lens plane at
$\left(x_i,y_i\right)$ will experience a deflection due to the
microlensing masses and will intercept the source plane at a position
$\left(x_s,y_s\right)$. Considering a single source, at position
$\left(x_R,y_R\right)$, it can be seen that once a grid of $N_{r}$
rays has been traced through the lensing plane, the centroid of the
macroimage, due to the individual microimages, is at;
\begin{equation}
\left< x \right> = \frac{1}{N_{r}} {
\sum_{\scriptscriptstyle{N_{r}}} x_i \ \
\rho\left(x_s,y_s|x_R,y_R\right)}\ \ ,
\end{equation}
with a similar equation for the y co-ordinate. Here, the function
$\rho$ describes the surface brightness distribution of the source.
For the study presented here, a source is assumed to be a disk of
radius ${\rm R}$, with a uniform surface brightness distribution. More
general source profiles can be employed, although for this study the
scale of any centroid shift is of interest and this will not depend
heavily on the detail of the source profile.

\subsection{Model Parameters}\label{microlensingmodel_parameters}

As with the earlier work of Williams and Saha (1995), we consider
image centroiding in the multiply imaged quasar, Q2237+0305.  Since
its discovery, a number of models, both parametric and non-parametric,
for the lensing mass distribution in this system have been
developed. Using ground--based (Kent and Falco 1988, Schneider et al
1988), and later, Hubble space telescope data (Crane et al 1991, Rix
et al 1992, Wambsganss and Paczy\'{n}ski 1994, Schmidt et al 1998),
these models can, to within the observational error, reproduce the
image configuration.  Many models, however, fail to accurately
reproduce the observed optical flux ratios, although there is broader
agreement with the ratios from the larger radio emitting region (Falco
et al. 1997). It has been suggested that this discrepancy is due to
the action of differential microlensing, with the smaller UV--optical
emitting continuum source being more susceptible to the induced
fluctuations (Schmidt et al 1998), although the effects of
differential reddening between the images has also been considered
(Nadeau et al. 1991).

Although these models vary greatly in the details, those possessing
parameterized, elliptical forms for the mass distribution, without the
need for an additional external source of shear, tend to predict
similar optical depth and shear parameters at the positions of the
images~\footnote{It should be noted that this is also the case for the
brightest of the three ``mini-images'' that comprise image B of the
non-parametric model of Williams and Saha (1995).}. For simplicity we
employ the ``best--fit'' model of Rix et al. (1992), which employs an
elliptical $r^{-\frac{1}{4}}$ mass distribution with an unresolved
core.  The microlensing parameters at the positions of the images are
summarized in Table 1~\footnote{There is an error in the equation for
the shear parameter given in Table 3 of Rix et al. (1992). This should
read $\gamma = \sqrt{ \left( 1 - \tau \right)^2 - 1/Ampl }$, where
$\tau$ is the microlensing optical depth $\left(\equiv\sigma_*\right)$
and $Ampl$ is the image amplification. Taking the Rix et al. (1992)
values for the optical depth and amplifications, the shear values at
the positions of the images should be $\gamma_A=0.43$,
$\gamma_B=0.43$, $\gamma_C=0.69$ and $\gamma_D=0.62$. The differences
between these and the employed microlensing parameters do not effect
the overall conclusions of this paper.}

\vbox{
\begin{center}
\begin{tabular}{|c|c|c|c|}
\hline
Image & $\sigma_*$ & $\gamma$ & $\mu_{th}$ \\ \hline
A     &  0.41      &  0.47    &   7.86     \\
B     &  0.38      &  0.43    &   5.01     \\
C     &  0.65      &  0.68    &   2.94     \\
D     &  0.59      &  0.56    &   6.87     \\ \hline
\end{tabular}
\end{center}
{{\sc Table~1.} The macrolensing parameters, the surface density of
matter, $\sigma_*$, and large scale shear, $\gamma$, at the positions
of the quasar images in Q2237+0305. The values are the ``best'' model
parameters of Rix et al. (1992), and $\mu_{th}$ is the ``macrolensing
amplification''.  The image designations are those of Yee (1988).
\label{parameter_table}}
}
 
\smallskip 

For each image, two simulations were performed.  Firstly, the
microlensing stars were distributed with a uniform mass function,
whereby all objects have a mass of 1${\rm M_{\odot}}$, and secondly
according to a Salpeter mass function of the form
\begin{equation}
p(m)~dm \propto m^{-2.35}~dm \,
\end{equation}
between the limits $0.05{\rm M_{\odot}} < {\rm M} < 10{\rm
M_{\odot}}$, with ${\rm \left<M\right> }= 0.163{\rm M_{\odot}}$.

Three source radii were employed, ${\rm R= 5\times10^{14}}$, ${\rm
5\times10^{15} }$ \& ${\rm
5\times10^{16}h_{75}^{\scriptscriptstyle{-\frac{1}{2}}} cm}$ which are
of order the scales of structure expected for continuum emission from
an accretion disk at a quasars core (Ress 1984). For these sources, a
total of 1275 rays per pixel were traced through the image plane,
corresponding to a total number of rays, collected in the source
plane, of 84 million per simulation.  To ensure that the vast majority
of micro--images are considered, the total number of rays traced per
simulation was several times greater than this value.  Considering the
smallest source, and assuming a maximum deamplification of 2
magnitudes, this number of rays ensure a maximum error of less than
10\% in the derived image amplification.

\section{Results}\label{results_section}

Figures~\ref{MSOLAR} and~\ref{SALPETER} present the results of this
study for the constant Solar mass distribution and Salpeter
distribution respectively.  Each figure consists of four main panels,
one for each image in Q2237+0305. Each panel presents, on the right
and left side, two trajectories across the amplification map, each
perpendicular to one another. With this, one trajectory is aligned
with the dominant component of the shear. The statistical properties
of induced variability a source experiences as it travels behind a
microlensing region depends on its orientation with respect to this
shear component, with more rapid variability occurring as the source
moves along the dominant shear component, although the small scale of
the source plane under consideration does not make this direction
immediately apparent.  Each Figure presents data for the three source
sizes, the black line representing the smallest source, the light grey
line, the largest.  The intermediate source is represented by a grey
line, although it follows the variability of the smallest source to a
high degree, with significant differences being apparent only in the
vicinity of caustics.

The top figures in each panel present the image light curves, showing,
in magnitudes, the fluctuations about the ``macrolensing
amplification'' (Table~\ref{parameter_table}). As expected, the
smallest source undergoes the most rapid and complex variability,
while the largest source exhibits long term trends.  Below each light
curve, two frames are presented. The first presents the degree of
centroid shift in the direction perpendicular to the shear. It should
be noted that this shift is expressed in units of 15 angular Einstein
radii. In all the frames this component is negligible on this scale,
with several showing low, long term trends. The bottom frames show the
centroid shift in the direction of the major shear component, and
these display considerable fluctuations on both short and long time
scales. As with the image light curves, these fluctuations are more
severe for the smallest sources, with the larger source following the
large scale trends in the variability. Any large change fluctuation in
the centroid position is correlated with a fluctuation in the total
image brightness, although the converse is not true and all
fluctuations in the image light curves do not result in large centroid
shifts. This indicates that large changes in the \macro\ centroid are
due to the production or disappearance of bright images during the
crossing of a caustic, although some crossings produce image pairs
close enough to a current image centroid to produce no shift.

To illustrate this effect in more detail we consider a small
displacement of the quasar source, and examine the resultant change in
the image configuration. Utilizing the simulation for Image A with all
stars possessing a mass of ${\rm 1~M_{\odot}}$ the source was placed
at two points straddling a caustic. These source positions are
illustrated, superimposed upon the caustic network, in
Figure~\ref{fig_microlens}. The position of the source, relative to
this caustic network, changes by only by a small amount ${(\rm \Delta
E_s = 0.2 ER)}$, but as seen in the upper--left panel of
Figure~\ref{MSOLAR}, the resultant change in the image position is
substantial, being ${\rm \Delta E_i \sim 3 ER}$ for the largest
source, and greater for the smaller sources, with ${\rm \Delta E_i
\sim 10 ER}$.  The distribution of images at these two source
positions are presented in Figure~\ref{Microimage}.  The `low'
resolution of this Figure allows us to consider in detail the effects
on the largest source only. At each of the time steps, the
distribution of \micros\ is dominated by three large (bright) images,
which lie along the direction of the dominant shear component.  If all
the matter in the galaxy were smoothly distributed the resultant image
centroid would be on the left of the frame (denoted by the $\odot$
symbol).  The smaller to larger circles also presented in the plot
denote the centroids of the smallest to largest source size under
consideration.  Between the frames, the left--hand images fade, as the
source moves from the region of high amplification, through the
caustic, and into the deamplification region. One of these images
(far--left) appears to be the pair of images merging due to the
caustic crossing. Due to its extent, the wing of the largest source
still overlaps the caustic at the latter time step and possess an
appreciable image on the far left of the frame. The smaller sources,
which would completely crossed the caustic by this point, have lost
these bright images.

The large--scale distribution of \micros\ is also apparent in Figure~3
of Wambsganss (1992). This presents the area of the image plane from
which rays are traced into a particular square region of the source
plane, as compared to a similar region which would map directly onto
the source plane area if the material in the microlensing stars was
smoothly distributed. The area of the image plane is effectively the
image of the square source plane region and is composed of five major
components, although, due to the small region of this image plane
under consideration, one of images extends beyond the figure.  It is
apparent that the centroid of these images will not coincide with that
of the image created by a smooth matter distribution by a few Einstein
radii. This frame represents a single snap--shot of possible \micro\
configurations and more extreme fluctuations are likely.

\subsection{Time Scales and Physical Shifts}\label{physical_section}

Although not indicative of the time scales of individual events, the
crossing time of an Einstein radius (Equation~\ref{timescale}) still
provides a temporal measure of microlensing activity. Following Kayser
et al. (1986), this crossing time for a Solar mass star in Q2237+0305
is
\begin{equation}
{\rm
\tau = \left( 1 + z_l \right) \frac{\eta_o}{v_l} \frac{D_{ol}}{D_{os}} \sim
\frac{8.7}{v_{600}} h_{75}^{\scriptscriptstyle{-\frac{1}{2}}}  yrs}\ ,
\label{timescale_yrs}
\end{equation}
where ${\rm z_l}$ is the redshift of the lensing galaxy, and ${\rm v_l
= 600 v_{600} km~s^{-1}}$ is its departure velocity from the Hubble
flow, perpendicular to the line-of-sight. Figures~\ref{MSOLAR}
and~\ref{SALPETER} reveal that the smaller sources experience centroid
shifts on much shorter time scales than this, with substantial shifts
occurring on $\lta$yr periods, superimposed on longer term variability.

At a redshift of $z=0.04$, the lensing galaxy in Q2237+0305 lies at a
distance of 147${\rm h_{75}^{\scriptscriptstyle{-1}}}$Mpc, and the
angle subtended by the Einstein radius of a Solar mass star ${\rm (
0.0052h_{75}^{\scriptscriptstyle{-\frac{1}{2}}}pc)}$ is ${\rm
7.3\times10^{-6}h_{75}^{\scriptscriptstyle{\frac{1}{2}}}\scmd}$.  This
implies that the unit of centroid shift in Figures~\ref{MSOLAR}
and~\ref{SALPETER} is ${\rm
110\times10^{-6}h_{75}^{\scriptscriptstyle{\frac{1}{2}}}\scmd}$.

Considering current limits of optical astrometry, such centroid shifts
appear to be well below detection. In the next Section, however, we
consider the astrometric sensitivity of a forthcoming space craft,
the Space Interferometry mission.

\section{Space-borne Interferometry}\label{space_section}

Over its history, much of astronomy has focused upon the measurement
of distances to objects in the universe. Other than a small number of
``direct'' methods, such as gravitational lens time-delay measurements
(eg Kundi\'{c} et al. 1997, Schechter et al. 1997) or the detection of
a cluster's Sunyaev--Zeldovich decrement (eg Myers et al. 1997), most
rely on a ``ladder'' of techniques to determine the distance to
extra--Solar objects. The lowest rung of this ladder requires the
determination of distance to nearby objects, allowing a calibration of
their physical properties, and this can be achieved with a purely
geometrical measure, the parallax.  This method does, however, require
accurate determination of the ``shift'' of objects with respect to the
extra--galactic background as the Earth orbits the sun. The effects of
atmospheric seeing limits such measures, allowing a parallax
determination of fractions of an arcsecond.

Astrometric measurements will be revolutionized with the development
of space-borne optical interferometric techniques (Loiseau and Malbet
1995). Due to fly in 2005, the Space Interferometry Mission (SIM)
promises to surpass all previous space--borne astrometric missions
[e.g. Hipparcos (Perryman et al. 1992)] by several orders of
magnitude. Its current operational parameters will provide centroiding
of a $20^{\rm th}$ magnitude stellar image to $4\times10^{-6}\scmd$
with 14 hours of integration~\footnote{For a more comprehensive review
of the Space Interferometry Mission see {\tt
http://huey.jpl.nasa.gov/sim/}}. Such accuracy will provide parallax
determinations of Cepheids out to $\sim5$kpc, and will allow the
detection of `wobbles' in stellar proper motions due the presence of
Earth mass planets about nearby stars.  Recently, Miyamoto and Yoshii
(1995) proposed to use SIM to observe the $\sim20\mu\scmd$ shifts
expected during a microlensing event in the galactic halo, providing
more constraints on the lensing geometry.  As halo microlensing events
are typically due to an isloated MACHO, astrometric observations with
SIM will determine the mass, distance and proper motion of the
microlensing body, even if the MACHO is too faint to be detected
directly (Paczy\'{n}ski 1998).

It is immediately apparent that the scale of microlensed induced image
shifts in the centroids of macrolensed quasars $(m_R\sim 17-18.5)$ are
of sufficient magnitude and occur on short enough time scales to place
then are well within the operational capabilities of SIM.  As with
photometric monitoring, however, the complex nature of the caustic
distribution at high optical depths will prevent the association of a
particular image shifting ``event'' with an individual mass
(eg. Wambsganss 1992); rather the statistical properties of a number
of image shifts, such as their duration and degree, will provide
measure of a characteristic lensing mass and scale--size for the
continuum emitting region in the core of the quasar.  Importantly, the
dominant direction of image shifting is coincident with the direction
of the dominant shear component, and the determination of any shifts
perpendicular to this direction will provide a measure of the relative
values and orientation of the two amplification eigenvalues.  A
non--detection of any perpendicular shifts still reveals the
orientation of the amplification eigenvalues, but provides a limit on
their relative values. The identification of these components will
provide powerful additional constraints on any macrolensing mass model
for this system, leading to a more robust measure of the microlensing
optical depth and shear at the position of the images. 

\section{Conclusions}\label{conclusions_section}

This paper has investigated the r\^{o}le of microlensing in inducing
image shifts in the centroids of the star--like images of multiply
imaged quasars. It was found that for the quadruple quasar system,
Q2237+0305, that substantial shift of $\sim 100\mu\scmd$ are possible
on time scales of months. Although not observable with current
telescopes, such shifts will be detectable in the next decade with the
proposed Space Interferometry Mission.  This initial study has
revealed that the characteristics of centroid shifting are sensitive
to the size of the source undergoing microlensing, providing an
independent probe of the scales of structure in the continuum emitting
region of quasars. Similarly, these characteristics should depend on
the mass function of the microlensing objects, although a proper
determination of any influence will require statistical study of an
ensemble of simulations.

We wish to emphasize, however, that the light curves and image shifts
presented here are drawn from only single realizations of the
microlensing in the images of Q2237+0305.  The results must,
therefore, be taken as purely illustrative of the effect.  Currently,
an ensemble of simulations are underway; once complete, these will be
employed in a study of the statistical properties of image centroid
shifting and their dependence upon the scale of the source and mass
function of the microlensing bodies.

\section{Acknowledgments}

L.~L~ R.~Williams is thanked for providing a copy of her microlensing
code, on which this study was based. We also thank D.~Peterson for
engaging discussions on the operational capabilities of SIM. Z.~Kuncic
is thanked for proof reading the manuscript and the anonymous referee
is thanked for constructive comments.

\newpage

\begin{figure}
\centerline{
}
\caption[]{
An amplification map for Image A in Q2237+0305
$\left(\sigma_*=0.41,\gamma=0.47\right)$.  The grey--scale is a
measure of (de)amplification and is in units of magnitudes. Note that
a ${\rm \Delta M}$ is with--respect--to the brightness the
macro--image would appear if all the matter was smoothly distributed.
The map has an extent of 2.5 Einstein radii for a Solar mass star
along each axis, and the microlensing stars were distributed with a
uniform mass of $1{\rm M_{\odot}}$. The circles represent two 
positions of the source which are discussed in more detail in 
Section~\ref{results_section}.
}
\label{fig_microlens}
\end{figure}

\begin{figure}
\centerline{
}
\caption[]{
The results for the individual images of Q2237+0305. All
the microlensing stars possessed a mass of ${\rm 1M_{\odot}}$. For
each of the images, (A,B,C,D), two sets of panels are presented, each
representing source trajectories that are either aligned or
perpendicular to the direction of the dominant shear component. Each
plot is a function of the source position along this trajectory in
units of Einstein radii for a Solar mass star in the source plane.
For each orientation, three plots are given. The top panels for each
image presented the light curve (in magnitudes) for the smallest
source (darkest line) through to the largest (lightest line). The
middle panel presents the centroid shift in the direction
perpendicular to the major shear axes, while the lower panels are the
centroid shift in the direction of the major shear component. Note
that in these lower plots that shift is expressed in units of 15
angular Einstein radii for a Solar mass star.  
}
\label{MSOLAR}
\end{figure}

\begin{figure}
\centerline{
}
\caption[]{
As for Figure~\ref{MSOLAR}, but for the stars distributed with a
Salpeter mass function. All the axes are in the same units as
Figure~\ref{MSOLAR}, including the `x--axis' which represents angular
Einstein radii of a Solar mass star.
}
\label{SALPETER}
\end{figure}
 
\begin{figure}
\centerline{
}
\caption[]{
The distribution of \micros\ for the two source positions presented in
Figure~\ref{fig_microlens}.  The units along the axes are angular
Einstein radii of a Solar mass star.  The dark grey areas are the
images for the largest source ${\rm
(R=5\times10^{16}h_{75}^{\scriptscriptstyle{-\frac{1}{2}}} cm)}$,
while the lighter grey area represents the intermediate source ${\rm
(R=5\times10^{15}h_{75}^{\scriptscriptstyle{-\frac{1}{2}}} cm)}$.
Note that the resolution of this image does not reveal the full image
structure for this smaller source, and for the ${\rm
R=5\times10^{14}h_{75}^{\scriptscriptstyle{-\frac{1}{2}}} cm}$
source. The $\odot$ symbol indicates the position of the center of the
image if all the lensing matter was smoothly distributed. Each circle
corresponds to the centroid of the \micro\ brightness distribution for
each source, the smallest circle being for the smallest source etc.
}
\label{Microimage}
\end{figure}


\newpage

\begin{references}

\reference{alcock1993}
	Alcock, C. et al., 1993, \nat\, 365, 621

\reference{barnes1986}
	Barnes, J. \& Hut, J., 1986, \nat\, 324, 446

\reference{chang1979}
	Chang, K. \& Refsdal, S., 1979, \nat\, 282, 561

\reference{chang1984}
	Chang, K., 1984, \aap\, 120, 157

\reference{corrigan1991}
	Corrigan, R. T. et al., 1991, \aj\, 102, 34

\reference{crane1991}
	Crane, P. et al., 1991, \apj\, 369, 59

\reference{falco1996}
	Falco, E. E., Perley, R. A., Wambsganss, J. \& Gorenstein, M. V.,
	1996, \aj\, 112, 897


\reference{huchra1985}
	Huchra, J. et al., 1985, \aj\, 90, 691

\reference{irwin1989}
	Irwin, M. J., Webster, R. L., Hewett, P. C., Corrigan, R. C. \&
	Jedrzejewski, R. I., 1989, \aj\, 98, 1989 

\reference{kayser1986} 
	Kayser, R., Refsdal, S. \& Stabell, R., 1986,
	\aap\, 166, 36

\reference{kent1988}
	Kent, S. M. \& Falco, E. E., 1988, \aj\, 96, 1570

\reference{kundic1997}
	Kundi\'{c}, T. et al., 1997, \apj, 482, 75

\reference{lewis1993}
	Lewis, G. F., Miralda-Escud\'{e}, J., Richardson, D. C. \&
	Wambsganss, J., 1993, \mnras\, 261, 647

\reference{lewis1995}
	Lewis, G. F. \& Irwin, M. J., 1995, \mnras\, 276, 103

\reference{lewis1996}
	Lewis, G. F. \& Irwin, M. J., 1996, \mnras\, 283, 225

\reference{lewis1997} 
	Lewis, G. F., Irwin, M. J., Hewett, P. C. \& Foltz, C.B., 1997,
	\mnras\, Accepted.

\reference{loiseau1995}
	Loiseau, S. \& Malbert, F., 1996, \aaps\, 116, 373

\reference{miyamoto1995}
	Miyamoto, M. \& Yoshii, Y., 1995, \aj\, 110, 1427

\reference{myers1997}
	Myers, S. T., Baker, J. E., Readhead, A. C. S., Leitch, E. M. \&
        Herbig, T., 1997, \apj\, 485, 1

\reference{nadeau1991}
	Nadeau, D. et al., 1991, \apj, 376, 430

\reference{ostensen1996}
	{\O}stensen, R. et al., 1996, \aap\, 309, 590

\reference{paczynski1986}
	Paczy\'{n}ski, B., 1986, \apj\, 301, 503

\reference{paczynski1998}
	Paczy\'{n}ski, B., 1998, {\it astro-ph/9708155}

\reference{perryman1992}
	Perryman, M. A. C. et al., 1992, \aap\, 258, 1

\reference{rees1984}
	Rees, M. J., 1984, \araa\, 22, 471

\reference{rix1992} Rix, H.-W., Schneider, D. P. \& Bachall, J. N., 1992,
	 \aj\, 104, 959

\reference{saust1994}
	Saust, A. B., 1994, \aaps\, 103, 33

\reference{schechter1997}
	Schecter, P. L. et al., 1997, \apj\, 475, 85

\reference{schmidt1997} 
	Schmidt, R., Webster, R. L. \& Lewis, G. F., 1997, \mnras\, Accepted

\reference{schneider1988}
	Schneider, D. P. et al., 1988, \apj\, 294, 66

\reference{schneider1992}
	Schneider, P., Ehlers, J. \& Falco, E. E., 1992, Gravitational Lenses,
	Springer-Verlag Press.

\reference{walsh1979}
	Walsh, D., Carswell, R. F. \& Weymann, R. J., \nat\, 279, 381

\reference{wambsganss1990}
	Wambsganss, J., Paczy\'{n}ski, B. \& Schneider, P., 1990,
	\apj\, 358, L3

\reference{wambsganss1992}
	Wambsganss, J., 1992, \apj\, 392, 424

\reference{wambsganss1994}
	Wambsganss, J. \& Paczy\'{n}ski, 1994,
	\aj\, 108, 1156


\reference{williams1995}
	Williams, L. L. R. \& Saha, P., 1995, \aj\, 110, 1471

\reference{witt1993}
	Witt, H.-J., 1993, \apj\, 403, 530

\reference{yee88}
	Yee, H. K. C., 1988, \aj\, 95, 1331

\reference{young1981}
	Young, P., 1981, \apj\, 244, 756

\end{references}
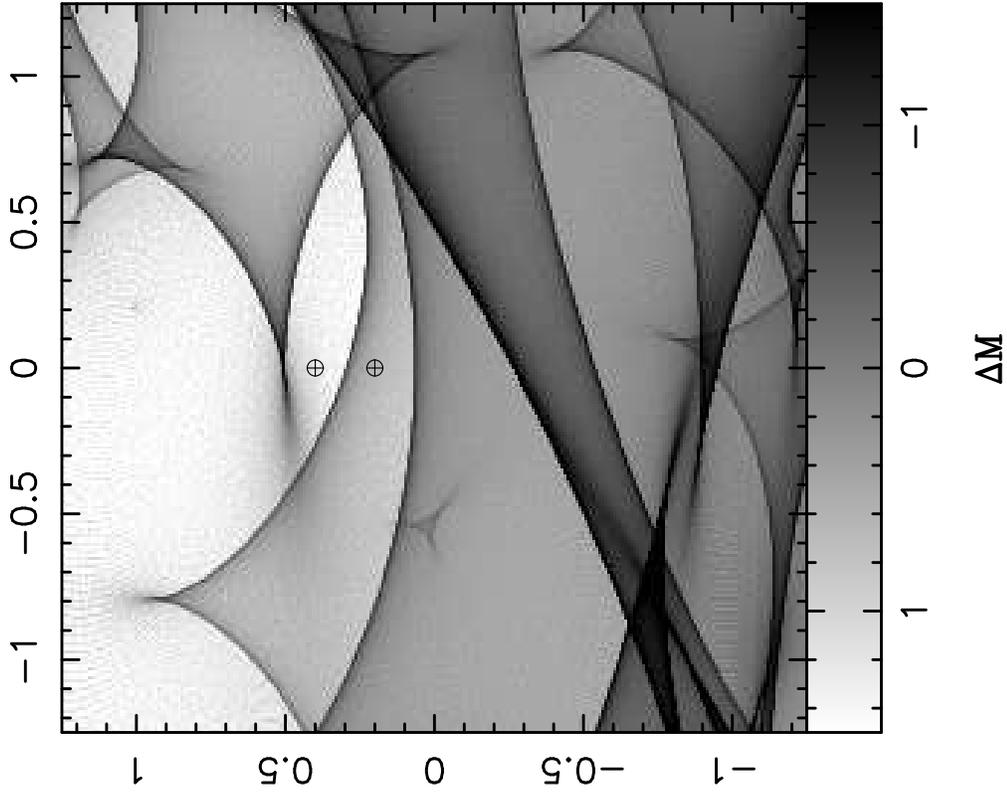
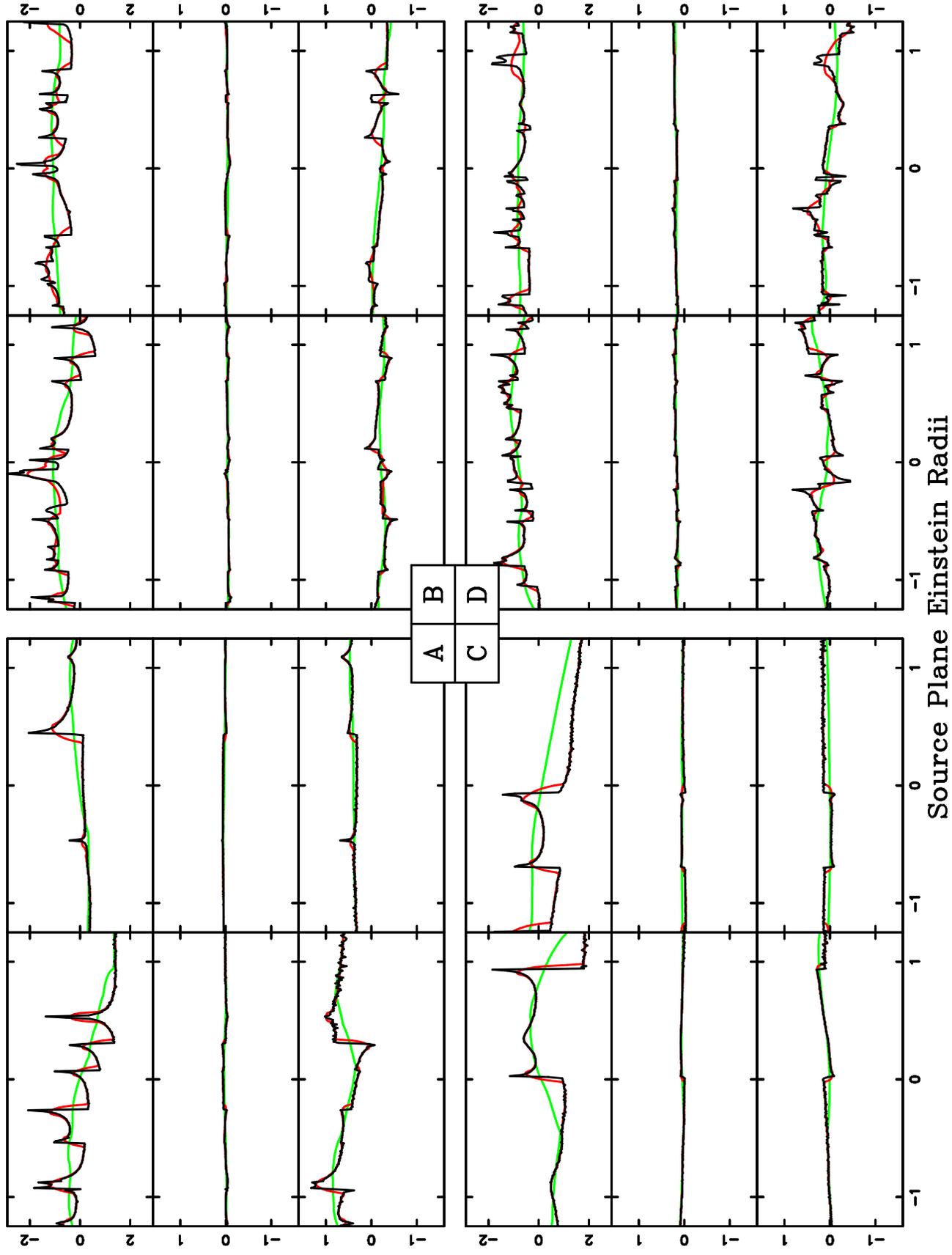
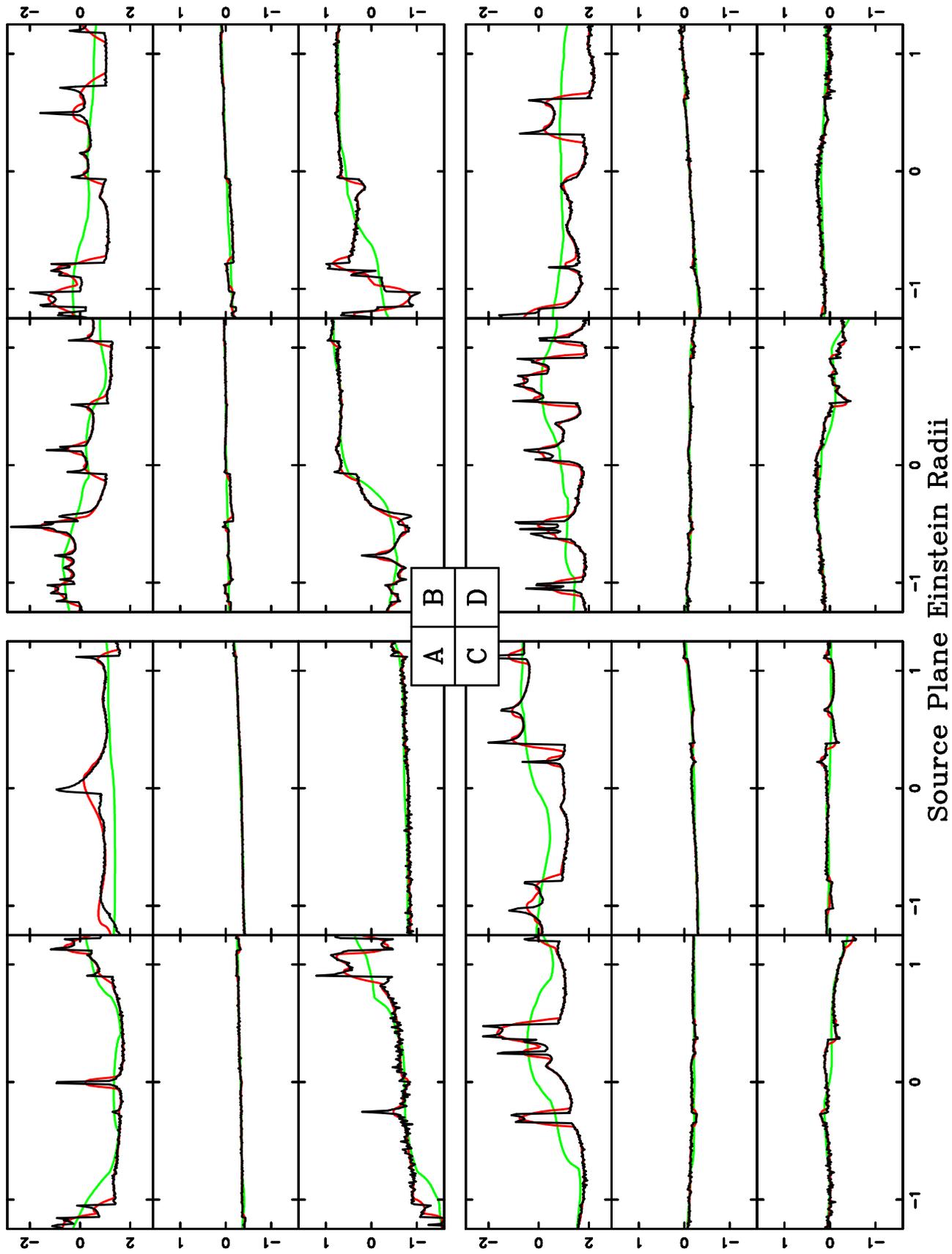
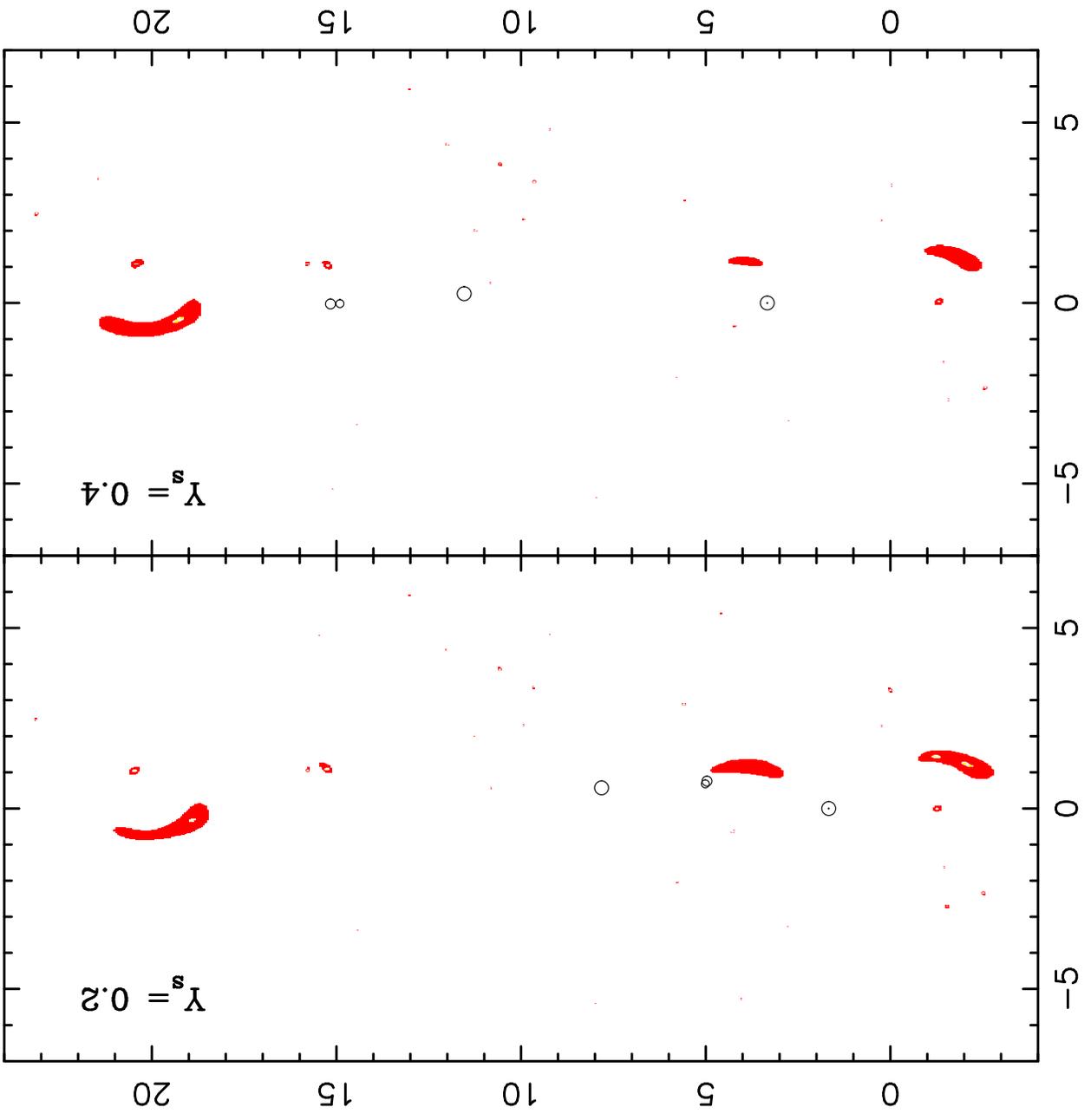
\end{document}